\documentstyle[prl,preprint,aps]{revtex}

\begin{document}
\draft
\preprint{ADP-93-202/M16}
\title{
Inconsistency between alternative approaches to \\
Quantum Decoherence in special systems
}
\author{J. Twamley}
\address{
Department of Physics and Mathematical Physics, University of Adelaide\\
GPO Box 498, Adelaide, South Australia 5001}
\date{\today}
\maketitle
\begin{abstract}
We study the decoherence properties of a certain class of Markovian quantum
open  systems from both the Decohering Histories and Environment
Induced Superselection paradigms. The class studied includes many
familiar quantum optical cases. For this class, we show that there
always exists a basis which leads to {\em
exactly} consistent histories for any coarse graining
{\em irrespective} of the initial conditions.
The magnitude of the off--diagonal elements of the reduced density
matrix $\rho$
in this  basis however, depends on the initial conditions.
Necessary requirements for
classicality as advanced by the two paradigms
are thus in direct conflict in 	these systems.
\end{abstract}
\narrowtext

In this letter we describe a class of quantum open Markovian
systems and examine the decoherence properties from the Decohering
Histories approach (DH) \cite{HARTLE:1991} and from the Environment Induced
Superselection approach (EIS) \cite{ZUREK:1991}. We curiously discover that for
this class of systems, histories defined as projections onto a
particular basis
($|n\rangle$ say) are {\em exactly}
consistent for any coarse graining
{\em irrespective} of the initial condition of the system.
Thus one can ascribe separate histories with probabilities which
obey the classical probability laws. In the framework of Decohering
Histories the attainment of a consistent set of alternative
histories is a
necessary requirement for the existence of a
``Quasiclassical Domain'' \cite{GELL-MANN:1990}.
In this same class of systems, however, the magnitude of the off--diagonal
elements of the reduced density matrix for the system {\em does} depend on
the initial state. Although the reduced density matrix
becomes diagonal in the basis $|n\rangle$ for times greater than
the dissipation
timescale, the  ``decoherence'' timescale in EIS
sensitively depends  on the initial
state of the system \cite{ZUREK:1991}. In EIS the criteria for classicality is
uncertain but generally a system is said to behave in a
classical manner when the reduced density matrix is sufficiently
diagonal in the ``pointer basis'' over a sufficiently long period of time
 (for the most recent report on EIS see \cite{ZUREK:1993}).
Thus, in these systems it appears that the requirements
for classicality in  EIS depends on the initial system state
whereas in DH  the requirements are independent
of the initial state. This result highlights the differences
between these two approaches and brings into question the validity of
either in describing the transition from  quantum to classical in these
models.

\paragraph{Models}
Consider the class of quantum systems which are linearly coupled to an infinite
bath of harmonic oscillators in the regime where the system's dynamics
are Markovian. For these  models one can write the master equation for
the reduced density matrix of the system in a standard form as shown
by Linblad \cite{LINBLAD:1976,GARDINER:QUANTUM_NOISE}
\begin{equation}
\dot{\rho}=L\rho=-\frac{i}{\hbar}[H,\rho]+
\sum_{J}[2A_{J}\rho A_{J}^{\dagger}-
\rho
A_{J}^{\dagger}A_{J}-A_{J}^{\dagger}A_{J}\rho]\;\;,\label{LINBLAD}
\end{equation}
where $H$ is some Hermitian operator and the $A_{J}$ are arbitrary. The
rigorous proofs require the $A_{J}$ to be bounded operators while a
number of authors have successfully applied Linblad's theory to
unbounded operators \cite{SANDULESCU:1987}.
We will concentrate only on those models where
the diagonal matrix elements of (\ref{LINBLAD}) in some basis
$|n\rangle$ involve only diagonal matrix elements of $\rho$. One such
model occurs when
the $A_{J}$'s are a representation of a Lie algebra $\cal L$
and where $H$ is
constructed solely from elements of the Cartan subalgebra $ h$ of
$\cal L$. Here $|n\rangle$ is given by the eigenbasis of
$A_{J}^{\dagger}A_{J}$ for some $J$ ie. an eigenbasis of an element of
the Cartan subgroup $ h$.
One can, however, construct other models where
again $|n\rangle$ and $H$  are as above while
the $A_{J}$'s can be complicated functions of single photon creation
and annihilation operators eg. $ A=a^{m}$, $A=f(a^{\dagger}a)$ or
$A= a+a^{\dagger}$. In all the above examples bar the last,
the basis $|n\rangle$ is discrete. For these types of models
the diagonal matrix
elements of (\ref{LINBLAD}) in $|n\rangle$ can be written as
\begin{equation}
\langle n|\dot{\rho}|n\rangle=\dot{\rho}_{nn}=\sum_{m=0}^{\infty}\,
c_{nm}\rho_{mm}\;\;.\label{DIAG_EVL}
\end{equation}
The dynamics of the diagonal elements of $\rho$ in this particular
basis decouples completely from the off--diagonal elements. We
effectively have a dynamical superselection rule. In fact, since the
diagonal elements are the probabilities, (\ref{DIAG_EVL}) may be
rewritten as
\begin{equation}
\partial_{t}\,P(n)=\sum_{m=0}^{\infty}\,c_{nm}P(m)\;\;,\label{BIRTH}
\end{equation}
which is a discrete version of a classical Markov process -- a birth
death stochastic process \cite{GARDINER:QUANTUM_NOISE}.

We now examine the Decoherence Functional for such  models
where we ignore the bath of harmonic oscillators and concentrate on
coarse grained histories of the system defined through projections
onto coarse grainings of $|n\rangle\langle n|$ at particular instants
of time. We can write the Decoherence Functional as
\begin{eqnarray}
&&{\bf D}(\alpha^{\prime},\alpha) = \nonumber\\
&&Tr\left[P^{\alpha_{i}}_{i}\left(\,e^{\int_{\Delta t_{i}}\,Ldt}
\left(P^{\alpha_{i-1}}_{i-1}\left(
e^{\int_{\Delta
t_{i-1}}\,Ldt}(\cdots)\right)P_{i-1}^{\alpha_{i-1}\,\prime}\right)
\right)\right]\;\;,\label{DF}\\
\end{eqnarray}
where we have taken the trace over the environment and used the master
equation (\ref{LINBLAD}) to evolve forward the reduced
density matrix of the system between subsequent projections $P_{i-2}$
and $P_{i-1}$ etc. We describe the histories through the $P^{\alpha}_{i}$
where
\begin{equation}
P^{\alpha}_{i}=\sum_{n\in n_{\alpha_{i}}}\,|n\rangle\langle
n|\;\;,\label{PROJ}
\end{equation}
where the $n_{\alpha}$ represents a complete and exclusive binning of
the basis $|n\rangle$ into alternatives labeled by $\alpha_{i}$.
Let us concentrate on the final $t_{i-1}$  to $t_{i}$ in the
Decoherence Functional (\ref{DF}). We have
\renewcommand{\arraystretch}{.5}
\begin{eqnarray}
&&{\bf D}(\alpha,\alpha^{\prime})=\nonumber\\
&&\sum_{\begin{array}{c}
n_{i}\in n_{\alpha_{i}}\\
n_{i-1}\in n_{\alpha_{i-1}}\\
n_{i-1}^{\prime}\in n_{\alpha_{i-1}^{\prime}}\end{array}}
\,\langle n_{i}|e^{\int_{\Delta
t_{i}}\,Ldt}
\left(\tilde{\rho}_{n_{i-1}\,n^{\prime}_{i-1}}|n_{i-1}\rangle\langle
n_{i-1}^{\prime} |\right)|n_{i}\rangle\;\;,\label{short_df}
\end{eqnarray}
where
\begin{equation}
\tilde{\rho}_{n_{i-1}\,n^{\prime}_{i-1}}=\langle n_{i-1}|
e^{\int_{\Delta
t_{i-1}}\,Ldt}(\cdots)|n^{\prime}_{i-1}\rangle\;\;.\end{equation}
However, from equation (\ref{DIAG_EVL}) the final trace in the
Decoherence Functional implies that the right hand side of
(\ref{short_df}) vanishes for
$n_{i-1}\neq n^{\prime}_{i-1}$. This argument can be repeated for times
$t_i$ and $t_{i-2}$ and so on. We ultimately find that the Decoherence
Functional exactly vanishes when the two histories $\alpha$ and
$\alpha^{\prime}$ differ.
Thus, for the types of models discussed above
the Decoherence Functional is {\em exactly} diagonal in the histories.
Further, from (\ref{short_df}) we see that the result
does not depend on the specific values of
$\tilde{\rho}_{n_{i-1}n_{i-1}^{\prime}}$. The achievement of
consistency is therefore
completely insensitive to the initial state of the system.
We also note that this result holds for all possible binnings occurring
in the construction of the projectors (\ref{PROJ}).

The dynamics of the off--diagonal elements of the reduced density
matrix of the system in the $|n\rangle$ basis is, of course, dependent on
the initial state. One can begin with an initial state where the
magnitude of the off--diagonal elements are very large. One example
where the off-diagonal elements decouple exactly from diagonal
elements is that of a spin with a magnetic moment $M$ in a fluctuating
magnetic field \cite{LINDENBERG_WEST:NONEQUILIBRIUM}.
Other examples can be found in
number of models commonly used in quantum optics. Specifically, we look at
master equations derived in the Markov-Born limit with weak coupling
to the bath and with the Rotating Wave Approximation (RWA). The master
equations for such quantum optical models are only approximate but are
generally regarded as good descriptions of the physical processes in the proper
regimes. We will have more to say concerning the validity of these
master equations later.

We take the case of a general system coupled to a thermal
bath and a broadband squeezed vacuum. The quantum optical master
equation is \cite{DUM:1992}
\begin{eqnarray}
\dot{\rho}_{\rm sys}=&& -\frac{i}{\hbar}[H_{\rm sys},\rho]
+\frac{1}{2}\gamma (N+1)(2c\rho
c^{\dagger}-c^{\dagger}c\rho-\rho c^{\dagger}c)\nonumber\\
&&+\frac{1}{2}\gamma N(2c^{\dagger}\rho c-cc^{\dagger}\rho-\rho
cc^{\dagger})\nonumber\\
&&-\frac{1}{2}\gamma M(2c^{\dagger}\rho
c^{\dagger}-c^{\dagger}c^{\dagger}\rho-\rho
c^{\dagger}c^{\dagger})\nonumber \\
&&-\frac{1}{2}\gamma M^{*}(2c \rho c - cc\rho -\rho
cc)\;\;,\label{big-master}
\end{eqnarray}
where $\gamma$ is the coupling strength to the bath, $c$ is a system
operator which effectively couples to the creation operator of the
bath (see \cite{GARDINER:QUANTUM_NOISE}), $N$ is the number of quantum
per mode in the reservoir and $M$ with $N(N+1)\geq |M|^2$ is a measure
of the squeezing.

Looking first at the case where the system is coupled to a thermal
vacuum, i.e. $N=0$, $M=0$, the types of system Hamiltonians which
match the models described in this note depend on the type of system
operator which couples to the bath. The simplest is where $c=a$, the
system lowering operator. The master equation is already in the
Linblad form where $|n\rangle$ is the eigenbasis of $a^{\dagger}a$.
We can take $H_{\rm sys}=f(a^{\dagger}a)$ where $f$ is arbitrary.
In particular we can set
$H_{\rm sys}=\hbar \omega(a^{\dagger}a+1/2)$, the simple
harmonic oscillator. We can also set $c=a^{\dagger}a$ with $|n\rangle$
and $H$ as before  to obtain the phase damped harmonic
oscillator. The decay of off-diagonal coherences for these examples
has been studied by Walls and Milburn \cite{WALLS:1985}.

One can include a thermal bath and still retain the diagonal property
of the master equation. Note however, that the inclusion of a driving field
destroys this property. Thus, histories in the photon number basis
will be automatically consistent in these models while the
off-diagonal elements in the number basis of $\rho$ may be quite large
over periods much greater than the Markov time.

In considering the additional coupling to a squeezed vacuum we follow
\cite{DUM:1992} and rewrite (\ref{big-master}) as
\begin{equation}
\dot{\rho}=-\frac{i}{\hbar}[H_{\rm
sys},\rho]+\Lambda\rho\;\;.
\end{equation}
We may recast $\Lambda\rho$ in the Linblad form,
\begin{equation}
\Lambda \rho=\frac{1}{2}\sum_{\kappa=1}^{2}\,\lambda_{\kappa}(2N+1)
[2a_{\kappa}\rho
a_{\kappa}^{\dagger}-a_{\kappa}^{\dagger}a_{\kappa}\rho
-\rho a_{\kappa}^{\dagger}a_{\kappa}]\;\;,
\end{equation}
where
\begin{equation}
a_{\kappa}=\sum_{\kappa=1}^{2}\,c_{i}V_{i\kappa}\qquad\qquad(\kappa=1,2)\;\;,
\end{equation}\begin{equation}
\lambda_{1,2}=\frac{\gamma}{2}(2N+1\pm\sqrt{1+4|M|^2})\;\;,
\end{equation}
\renewcommand{\arraystretch}{1}
\begin{equation}
V=\left(\begin{array}{cc}
\cos\, \frac{\theta}{2}\; e^{-i\phi/2} & -\sin\,
\frac{\theta}{2}\;e^{-i\phi/2} \\
\sin\, \frac{\theta}{2}\;e^{i\phi/2} &
\cos\,\frac{\theta}{2}e^{i\phi/2} \end{array}\right)\;\;,
\end{equation}
and $c_{1}\equiv c$, $c_{2}\equiv c^{\dagger}$.
We only consider the pure state squeezed vacuum case where $N(N+1)=|M|^2$.
In this case $\lambda_{1}=\gamma(2N+1)$, $\lambda_{2}=0$ with
\begin{equation}
a_{1}=(\sqrt{N+1}e^{-i\phi/2}c+\sqrt{N}e^{i\phi/2}c^{\dagger})\;\;,
\label{a-sol}\end{equation}
and $a_{2}=0$. We can show $[a_{1},a_{1}^{\dagger}]=1$ and thus
$\Lambda\rho$ will only couple diagonal elements of $\rho$ in the
$a_{1}^{\dagger}a_{1}$ eigenbasis. For the complete master equation to
couple only diagonal matrix elements we must have $H_{\rm
sys}=f\,[a_{1}^{\dagger}a_{1}]$ or
\begin{equation}
H_{\rm
sys}=f\,[(2N+1)a^{\dagger}a+\sqrt{N(N+1)}e^{i\phi}a^{\dagger\,2}
+Ne^{-i\phi}a^{2}] \;\;,\label{squeeze-cartan}\end{equation}
where $f$ is again an arbitrary function.
Although such a system Hamiltonian is quite artificial it is curious
that such a model yields consistent histories which obey classical
probability laws. The addition of $a^2$ and $a^{\dagger\,2}$ to the
system Hamiltonian usually results in the phenomena of ``squeezing''
\cite{GARDINER:QUANTUM_NOISE} and can produce very non--classical states.

\paragraph{Discussion}
We have described a class of models consisting of
system+interaction+bath  for which the definitions of ``decoherence''
in the Decohering Histories approach and Environment Induced
Superselection approach disagree. We find that there is an effective
dynamical superselection which decouples the diagonal matrix elements
of the reduced density matrix from the off-diagonal elements in a
particular basis.

Before discussing the possible implications of this
result let us comment on the validity of the quantum optical master
equation (\ref{big-master}) and in particular the Rotating Wave
Approximation (RWA). The original coupling of the system to the bath in
these models is taken to be of the position--position type ie.
$H_{\rm int}\sim
(a^{\dagger}+a)(b^{\dagger}+b)$ where the $a$ and $b$ are destruction
operators for the system and bath respectively. One then
neglects the high frequency components of this interaction,
$a^{\dagger}b^{\dagger}$ and $ab$ to finally obtain an interaction of
the form $a^{\dagger}b+ab^{\dagger}$. This procedure is almost
standard in all quantum optical calculations. However, as pointed out
by Lindenberg and West \cite{LINDENBERG_WEST:NONEQUILIBRIUM},
the fully--coupled model
(ie. without the RWA) possesses couplings between the diagonal and
off-diagonal elements of $\rho$ in the number basis. These couplings
are weak in the appropriate regimes and can usually be neglected but
their existence, no matter how small, eliminates such
position--position couplings from
the class of models studied in this paper. Interactions
between the system and the bath which do fall into the class of models
described in this note are those which involve linear couplings in
{\em both} position--position (x-x) and momentum--momentum (p-p).
These types of system--bath interactions can be
written in the $a^{\dagger}b+ab^{\dagger}$ form without any
approximation for any mixture of x-x and p-p coupling using a
canonical transformation. This dual interaction in both position and
momentum has received little attention in the literature. It has been
treated explicitly by Leggett \cite{LEGGET:1984} with a number of examples
of pure x-x (``normal''), pure (p-p) (``anomalous'') and ``mixed'' x-x
and p-p coupling. He concludes that in any realistic physical system the
dissipation is unlikely to be pure ``anomalous'' but can be of the
``mixed'' type. Mixed x-x and p-p couplings in Josephson junctions and quantum
tunnelling have also been treated in \cite{LOSS_MULLEN:1991,FORD_LEWIS:1991}.
Momentum dependent
interactions also occur frequently in nuclear physics, for example in meson
interactions \cite{CARINENA:1983,ATNAN:1985} and in
\cite{GARCIA-RECIO:1992}. From the framework
of Environment Induced Superselection we might also
expect a mixed x-x and p-p interaction. We normally picture classical
mechanics as existing on phase space. If we wish the correlations
between x and p for a quantum system coupled to a bath to exhibit
little or no interference on phase space, then
according to Zurek, one should couple the system to the bath in
both position and momentum.

It is clear that for the class of systems treated here the two
definitions of decoherence differ greatly. Further, in the DH
approach, the $|n\rangle$ projected histories are exactly consistent
irrespective of the initial state of the system for all grainings. If
one chooses to accept that this class of models can represent physically
realistic situations, (as we argue above)
then one is forced to conclude that either one or both of
the ``decoherence'' paradigms (EIS and DH) is incorrect as a sole
indicator of ``classical'' behaviour.
In DH the criteria for a quasi--classical world requires, at
least, one exactly consistent set of histories. This is achieved
almost automatically for all times greater than the Markov time
in these models. However, in EIS we can only say that the ``collapse''
of the wavefunction to the ``observed''
basis ($|n\rangle$) occurs for times greater than
the decoherence time $t_{d}$. This decoherence time depends on the
initial state of the system and can be of the order of
the relaxation timescale for sufficiently low temperatures and small
wavepacket spread. For very weak coupling $t_{d}$ may even exceed the age
of the universe! It has been argued that if the system
rapidly decoheres in the EIS approach for a particular basis then
coarse grained histories over this basis should be approximately
consistent \cite{ZUREK:1993}. An example where EIS decoherence is achieved
and DH decoherence is not has been discovered by Laflamme and Matacz
\cite{LAFLAMME:1993}.
We have found models displaying the converse.
Clearly, greater understanding of the physical implications of
these two paradigms is needed.

The author thanks Prof. W.H. Zurek and the T6 group at LANL for their
kind hospitality and insightful discussions during the period of this
research. We also thank Prof. G. Milburn and Dr. J. McCarthy for
stimulating discussions.

\end{document}